\documentclass[twocolumn,apj]{openjournal}
\usepackage{amsmath}

\newcommand{\perimeter}{Perimeter Institute for Theoretical Physics, 31 Caroline St N, Waterloo, ON N2L 2Y5, Canada}
\newcommand{\Berkeley}{Department of Physics, University of California, Berkeley, CA 94720, USA}

\graphicspath{ {plots/} }

\begin{document}

\title{Probing Dark Energy Microphysics with kSZ Tomography}

\author{Julius Adolff}
\email{juliusadolff@berkeley.edu}
\affiliation{\perimeter}
\affiliation{\Berkeley}

\author{Selim Hotinli}
\affiliation{\perimeter}

\author{Neal Dalal}
\affiliation{\perimeter}

\begin{abstract}
The accelerated expansion of the Universe is well established by geometric probes, yet its physical origin remains poorly understood. Most constraints on dark energy arise from background observables---supernovae, baryon acoustic oscillations, and the cosmic microwave background---which mainly test the homogeneous expansion history. To move beyond this limitation, we examine how kinetic Sunyaev--Zel'dovich (kSZ) tomography, combined with galaxy clustering, can probe perturbative effects of dark energy and improve constraints on its background parameters. Using a Fisher-matrix analysis of the joint power spectra for LSST- and CMB-S4-like surveys, we quantify the additional information kSZ tomography contributes to dark-energy inference. Including kSZ data tightens constraints on $w_0$ by 15 \% and on $w_a$ by 32 \%, with parameter degeneracies distinct from those of geometric probes. We also assess the detectability of dark-energy perturbations through a two-parameter model, finding that for canonical sound speed ($c_s=1$) the effects are sub-percent and confined to horizon scales, while smaller sound speeds shift them to accessible $k$-ranges. Near-term kSZ measurements will primarily serve to test the consistency between background and perturbative signals, while future low-noise, high-resolution surveys may begin to uncover the microphysical properties of dark energy.

\end{abstract}

\maketitle

\section{Introduction}
\label{sec:intro}
Over the past decades, the accelerated expansion of the Universe has been firmly established through multiple observational probes, including type Ia supernovae, baryon acoustic oscillations (BAO), and measurements of the cosmic microwave background (CMB). Within the standard cosmological paradigm, $\Lambda$CDM, this acceleration is attributed to a cosmological constant ($\Lambda$) representing vacuum energy. Despite its remarkable simplicity and empirical success, $\Lambda$CDM offers no compelling theoretical explanation for the observed value of the vacuum energy. In order to account for possible deviations from this minimal picture, many phenomenological extensions has been proposed. These models do not in themselves resolve the most fundamental questions about dark energy, but rather provide a framework for parameterizing possible departures from $\Lambda$CDM that are consistent with observations. A common approach is to expand the dark-energy equation of state as \cite{Linder_2003}
\begin{equation}
    w(a) = w_0 + (1-a)w_a,
\end{equation}
where the cosmological constant corresponds to $(w_0, w_a) = (-1,0)$.

Recent observational efforts focused on the geometric measurements of the smooth expansion history (i.e.~background observables). Most notably, the Dark Energy Spectroscopic Instrument (DESI)~\cite{DESI:2016fyo,
DESI:2022xcl} have reported potential deviations from $\Lambda$CDM at the $(2$--$4)\sigma$ level, depending on the choice of supernova dataset~\cite{deside25} included in the analysis. While these results remain tentative, they motivate a renewed focus on sharpening our empirical constraints. In particular, complementary probes of the cosmic expansion history---beyond supernovae, BAO, and CMB measurements---can help refine our knowledge of the background evolution and thus better probe the $(w_0,w_a)$ parameter space. However, background probes alone are unlikely to fully capture the phenomenology of dark energy. To distinguish among models and to approach a more fundamental understanding, it may be helpful to also investigate the role of perturbations. 

If dark energy is sourced from a dynamical field, then, like any field, dark energy should exhibit perturbations. These dark energy perturbations can be described, for example, in terms of fluctuations of a perfect fluid with equation of state $w=p/\rho$ and sound speed $c_s^2 = \delta p/\delta \rho$; or more generally within the Parameterized Post-Friedmann (PPF) framework \cite{Hu_2007}; which provides a model‑independent prescription that enforces consistency with the background expansion and energy–momentum conservation. As in the case of any field exhibiting perturbations, dark energy fluid would also cluster on scales beyond its sound horizon, leading to observational signatures that can in principle be detected. In the canonical quintessence model, where dark energy is represented by a scalar field that is minimally coupled to gravity with $c_s = 1$, for example, dark energy clusters only on horizon scales, whereas in models with reduced sound speed, dark energy can in principle cluster on all scales. Regardless of the specific model, dark-energy perturbations leave scale-dependent imprints on the matter power spectrum by modifying the growth of structure through gravitational interactions, producing enhancements or suppressions on large scales relative to the $\Lambda$CDM expectation. Probing these signatures offers a pathway to go beyond background-level constraints and directly test the \textit{microphysics} of dark energy.

In this regard, one particularly promising avenue is the use of the cosmological signatures induced by the interaction of CMB photons with the large-scale structure (CMB secondaries) such as the kinetic and thermal Sunyaev Zel'dovich effects, cosmic infrared background and integrated Sachs-Wolfe effects. The kinetic Sunyaev–Zel'dovich (kSZ) effect arises from the scattering of CMB photons off moving free electrons in large-scale structures \cite{Sunyaev:1980nv}.

Measurements of the kSZ effect can be utilized to reconstruct the large-scale bulk radial velocity field, providing a novel window into the growth of fluctuations. The joint analysis of these reconstructed velocity measurements together with the large-scale galaxy density distribution allow for precise measurements of small deviations from $\Lambda$CDM in the large-scale matter modes~\citep{Deutsch:2017ybc, Smith:2018bpn, Giri:2020pkk, Cayuso:2021ljq}. Recent theoretical work has demonstrated that such cross-correlations enable powerful sample-variance cancellation \cite{M_nchmeyer_2019, Hotinli:2019wdp, AnilKumar:2022flx, Kumar:2022bly}, greatly enhancing sensitivity to subtle effects such as those induced by dark-energy perturbations. These techniques have already been applied to data in Refs.~\citep{Hotinli:2025tul,Lai:2025qdw,Lague:2024czc,McCarthy:2024nik,Krywonos:2024mpb,Bloch:2024kpn} leading to high signal-to-noise detections of the galaxy-velocity cross-spectrum in Refs.~\citep{Hotinli:2025tul,Lai:2025qdw,Lague:2024czc} and yielded competitive constraints on primordial non-Gaussianity ($f_{\rm NL}$)~\cite{Hotinli:2025tul}.

In this paper, we explore the prospects of probing dark energy through a joint analysis of galaxy clustering and kSZ velocity reconstruction from upcoming surveys, with a focus on Large Synoptic Survey Telescope (LSST)~\citep{LSSTDarkEnergyScience:2012kar,2019ApJ...873..111I} and the CMB Stage-4 (CMB-S4)~\citep{CMB-S4:2016ple,Abazajian:2019eic} experiment. We highlight how the statistical power of galaxy and velocity power spectra, and their cross-correlations, adds to the suite of existing background probes. Specifically, we forecast the impact of dynamical dark energy---consistent with current BAO, CMB, and supernova data---on the matter power spectrum, and we quantify how LSST and CMB-S4 observations can tighten constraints on $(w_0,w_a)$ and break the degeneracies present in background-only analyses. By demonstrating how measurements of large-scale, scale-dependent clustering can contribute to constraining dark-energy phenomenology, we also assess the detectability of dark-energy perturbations in forthcoming data; emphasizing the potential of kSZ tomography to refine our cosmological models and to advance our understanding of the nature of dark energy.

This paper is organized as follows: In Sec. \ref{secII} we review the phenomenology of dark energy, including both background evolution and perturbations. Sec. \ref{secIII} summarizes current observational constraints from background probes and motivates the use of the matter power spectrum. In Sec. \ref{secIV} we describe our forecasting methodology, including survey specifications, the kSZ effect, and the Fisher matrix formalism. Sec. \ref{secV} presents the results of our forecasts, highlighting the role of kSZ tomography in constraining dark-energy parameters. We conclude in Sec. \ref{secVI} with a discussion of the implications and future prospects.

\section{Dark Energy Phenomenology}\label{secII}

The existence of dark energy was first established through observations of the expansion history of the universe. The second Friedmann equation,
\begin{equation}
    \frac{\ddot a}{a} = - \frac{4\pi G}{3}(\rho + 3p),
\end{equation}
implies that the observed accelerated expansion ($\ddot a > 0$) requires a dominant energy component that violates the strong energy condition ($\rho + 3p > 0$). A natural candidate is vacuum energy, which satisfies $\rho = -p$ and can therefore drive the observed acceleration. However, a wide range of alternative models has been proposed~\citep[see e.g.][for a recent comprehensive review]{Abdalla:2022yfr}, including $f(R)$ gravity~\citep{Sotiriou:2008rp,Nojiri:2010wj,DeFelice:2010aj,Capozziello:2011et,Clifton:2011jh} and quintessence~\citep{Ratra:1987rm,Wetterich:1994bg,Caldwell:1997ii,Zlatev:1998tr,Peebles:2002gy,Caldwell:2003vq,Carroll:2003wy,Copeland:2006wr}. A central goal is thus to establish observational tests capable of distinguishing among these possibilities and thereby deepening our understanding of dark energy.

In an unperturbed FRW universe, the energy-momentum tensor of any component (like dark energy) takes the form
\begin{equation}
T^0_0 = -\rho, \qquad T^i_j = p\,\delta^i_j.    
\end{equation}
Energy–momentum conservation ($\nabla_\mu T^{\mu\nu} = 0$) then uniquely determines the evolution of the field, provided its energy density and pressure are related by a known equation of state $w = p/\rho$. In the absence of compelling theoretical guidance, one often adopts a phenomenological parametrization, such as the linear expansion \cite{Linder_2003}
\begin{equation}\label{eqn:de_expension}
    w(a) = w_0 + (1 - a)w_a.
\end{equation}
Observations of the expansion rate $H(z)$ can then constrain $w_0$ and $w_a$. However, it is non-trivial to map these constraints directly onto fundamental dark energy models \cite{Shlivko:2024llw}. For example, quintessence models (which require $w > -1$) may nevertheless yield best-fit parameters $(w_0, w_a)$ that appear to cross the phantom divide \cite{Wolf_2024}. This typically occurs when $w$ only departs significantly from $-1$ at late times.

Measurements of $H(z)$ are obtained from a combination of CMB, supernova, and BAO data, and have already placed meaningful limits on $w_0$ and $w_a$. Recently, DESI has reported deviations from the $\Lambda$CDM prediction $(w_0, w_a) = (-1,0)$ at $> 3\sigma$ significance (depending on the supernova dataset employed) \cite{deside25}. This motivates considering alternative theories of dark energy and investigating their observational consequences. 

\subsection{Perturbations}

Whereas the background evolution of dark energy is, by symmetry, described by a single function $w(a)$, its perturbations are far less constrained. The dynamics of dark energy perturbations depend sensitively on the underlying mechanism driving acceleration. For example, modified gravity theories (such as $f(R)$) can predict perturbative behavior very different from quintessence, even when their background evolution is nearly identical \cite{Ishak_2006, Lue_2004}.

This theoretical uncertainty complicates the construction of a model-agnostic phenomenology but simultaneously presents an opportunity: perturbation-level observables can serve as powerful discriminants between competing dark-energy models. In this work, we adopt a fluid description of dark energy—or, when such a description becomes inadequate, its generalization within the Parameterized Post-Friedmann (PPF) framework \cite{Hu_2007}. We restrict our analysis to scales $k \lesssim 0.1\,\mathrm{Mpc}^{-1}$, where linear perturbation theory provides an accurate description \cite{Bernardeau_2002}. This approximation breaks down on very large scales, where general-relativistic effects become relevant \cite{Yoo_2010}, and on small scales, where non-linear matter clustering and mode coupling dominate \cite{Bernardeau_2002}. These limitations, however, have only a minor impact on our results: large-scale modes are observationally limited by cosmic variance and for non-vanishing sound speed, dark energy remains smooth on non-linear scales. The fluid description represents the minimal deviation from $\Lambda$CDM and is exact for quintessence models \cite{Hu_1998}. Should future observations reveal significant tensions with this picture, it may then be necessary to consider extensions beyond this framework.

Within the fluid picture, dark energy influences the matter power spectrum in two key ways. First, the accelerated expansion induces a decay of the gravitational potential $\Phi$, with a rate that depends on $w$ and $\Omega_m$ \cite{Ma_1999}. To illustrate this, recall that on super-horizon scales the spatial curvature,
\begin{equation}
    \zeta = \frac{5 + 3w}{3(1 + w)} \Phi\,,
\end{equation}
is conserved \cite{Baumann:2022mni}. As $w$ decreases from $0$ toward $-1$, $\Phi$ must decay as $(1 + w)/(5 + 3w) \approx \tfrac{1}{2}(1 + w)$ to preserve $\zeta$. Since matter overdensities evolve according to \cite{Ma_1995}
\begin{equation}
    \ddot \delta_m + 2 H \dot \delta_m = -k^2 \Phi,
\end{equation}
this decay of $\Phi$ suppresses the growth of $\delta_m$. Moreover, for fixed $\Omega_m$, changing $w$ shifts the onset of dark energy domination, altering not only the rate but also the timing of the decay of $\Phi$ \cite{Ma_1999}.

This background-driven effect is largely independent of the detailed microphysics of dark energy, and is typically the dominant influence on structure growth. Because it is scale-independent, precise small-scale measurements of the matter power spectrum can be exploited to constrain $(w_0, w_a)$ independently of how (or whether) dark energy clusters.

\begin{figure}
    \centering
    \includegraphics[width=1.0\linewidth]{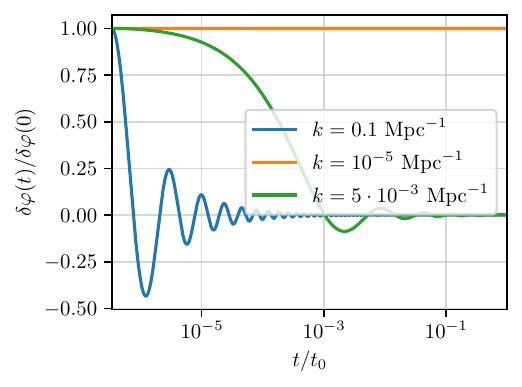}
    \caption{Evolution of quintessence perturbations $\delta \varphi$ for three representative scales. Modes quickly decay once they enter the sound horizon, while super-horizon modes remain frozen. The resulting transfer function is set by equation~\eqref{eq:Decay}.}
    \label{fig:Quint}
\end{figure}

There is, however, an important caveat: the conformal Newtonian gravitational potential $\Phi$ is not determined by the background alone, but rather by the full set of perturbations to the energy density \cite{MUKHANOV1992203}
\begin{equation}
    k^2 \Phi + 3 \mathcal H (\Phi' + \mathcal H \Phi) =  -4 \pi G\left(\rho_m \delta_m + \rho_{\rm de}\,\delta_{\rm de}\right).
    \label{eq:Poisson}
\end{equation}
If the dark energy perturbation $\delta \rho_{\rm de} = \rho_{\rm de}\delta_{\rm de}$ is non-negligible, it too can source $\Phi$ and thus modify the conclusions above, typically in a scale-dependent way. Addressing this requires specifying a model. In the fluid framework, $\delta \rho_{\rm de}$ is significant only outside the sound horizon of dark energy, determined by its sound speed $c_s$. Canonical quintessence predicts $c_s = 1$, pushing the sound horizon to horizon scales \cite{Creminelli_2009}. In this case, the previous conclusions hold, with  additional scale-dependent modifications appearing only on the largest observable scales.

We illustrate these points in the analytically tractable case of perturbations in a slowly varying quintessence field (with $\varphi_0 \approx \mathrm{const}$) during matter domination. In this regime, the perturbations $\delta \varphi$ satisfy \cite{Hu_1998}
\begin{equation}
    \ddot {\delta \varphi} + 3 H\dot{\delta \varphi} + \left((k/a)^2+V''(\varphi_0) \right)\delta \varphi= 2V'(\varphi_0)\Phi\,,
    \label{eq:Pert}
\end{equation}
with $H = 2/(3t)$, $a = (t/t_0)^{2/3}$, and $\Phi = \mathrm{const}$. On small scales, where $k^2 \gg a^2 V''(\varphi_0)$, Eq.~\eqref{eq:Pert} reduces to
\begin{equation}
        \ddot {\delta \varphi} + 3 H\dot{\delta \varphi} + (k/a)^2\delta \varphi=0. 
\end{equation}
The solution is
\begin{equation}
    \label{eq:Decay}
        \delta \varphi(t) \simeq \delta \varphi(0)\,\frac{\sin \left(2q\right)}{q^2} + \mathcal O\left(\frac{1}{q^3}\right), \quad q = \frac{k}{aH}. 
\end{equation}
Here, the normalization was fixed by requiring that modes decay once $k \sim aH$, while we neglected an overall $\mathcal{O}(1)$ coefficient and a possible phase shift.
On large scales, where $k^2 \ll a^2 V''(\varphi_0)$, Eq.~\eqref{eq:Pert} simplifies to
\begin{equation}
    \ddot{\delta \varphi} + 3 H \dot{\delta \varphi} + V''(\varphi_0)\,\delta \varphi
    = 2 V'(\varphi_0)\,\Phi,
\end{equation}
which has the solution (up to decaying modes)
\begin{equation}
    \delta \varphi = 2\Phi\,\frac{V'(\varphi_0)}{V''(\varphi_0)} = \mathrm{const}.
\end{equation}

The evolution of modes $\delta \varphi_k$ on different scales is shown in Fig.~\ref{fig:Quint}. Eq. \eqref{eq:Decay} in particular explains the $\sim 1/k^2$ dependence found in Ref.~\cite{Ma_1999} for the scale-dependent modification that quintessence induces in the matter power spectrum.

\begin{figure}
    \centering
    \includegraphics[width=1.0\linewidth]{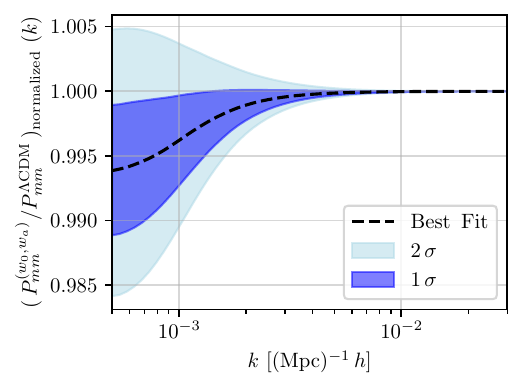}
    \caption{Scale-dependent modifications to the matter power spectrum induced by dark energy perturbations. We plot the ratio of the matter power spectrum in a $w_0w_a$ model to that of $\Lambda$CDM, normalized on small scales where dark energy perturbations are negligible. We plot this ratio for all values of $w_0w_a$ sampled from the DESI BAO, PantheonPlus SNe, and Planck CMB posterior. Best-fit values predict a $\sim 1\%$ effect on horizon scales. We used the PPF formalism as implemented in $\mathtt{camb}$ \cite{Lewis_2000, Fang_2008} to compute these and subsequent matter powerspectra.  }
    \label{fig:PowerSpecw0wa}
\end{figure}

We therefore conclude that dynamical dark energy with non-vanishing perturbations modifies the $\Lambda$CDM matter power spectrum in two ways. First, the altered background evolution changes the decay of the gravitational potential $\Phi$, leading to scale-independent modifications in the growth rate of matter perturbations. Second, on scales larger than the dark energy sound horizon, dark energy perturbations themselves source $\Phi$ according to Eq.~\eqref{eq:Poisson}. If the sound speed is canonical ($c_s = 1$), the sound horizon is of order the Hubble scale, $k_s \sim aH$, and this second effect is small, producing only a nearly unobservable correction on horizon scales. The impact of this effect is illustrated in Fig.~\ref{fig:PowerSpecw0wa}, which shows the scale-dependent modifications to the power spectrum derived from the DESI $w_0w_a$ posterior.

\pagebreak
\section{Observables}\label{secIII}

\subsection{Current Constraints from Background Measurements}

\begin{figure}
    \centering
    \includegraphics[width=1.0\linewidth]{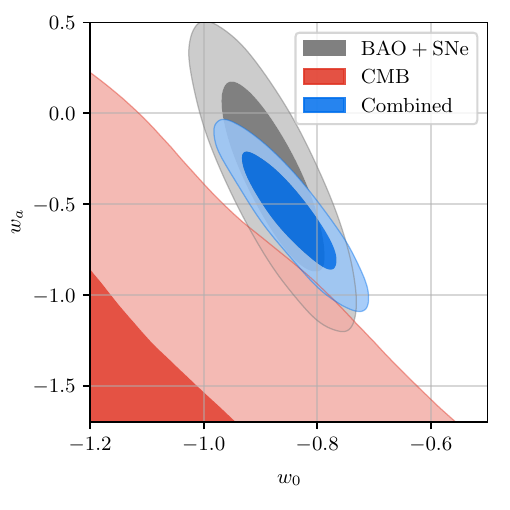}
    \caption{Constraints on $(w_0, w_a)$ from different background probes and their combinations (see text for details). For the BAO and SNe analyses, we impose a Gaussian BBN prior on the baryon density \cite{deside25} in order to break the degeneracy between baryonic and cold dark matter components. Here, and throughout, posteriors were sampled using the $\mathtt{cobaya}$ \cite{Torrado_2021, 2019ascl.soft10019T, 2021JCAP...05..057T} and $\mathtt{pocomc}$ \cite{karamanis2022accelerating, karamanis2022pocomc} python packages. The samples were analyzed using the $\mathtt{getdist}$ \cite{Lewis_2025} package. }
    \label{fig:Constraints1}
\end{figure}

The dark energy equation of state in Eq.~\ref{eqn:de_expension} is constrained using a variety of cosmological probes. Of particular importance are measurements from the CMB, BAO and SNe~\cite{deside25}.  

\paragraph{CMB constraints.}  
At the time of recombination, dark energy contributes only negligibly to the total energy budget of the Universe. As a result, primary CMB anisotropies are only weakly sensitive to $(w_0, w_a)$ and probe dark energy mainly through secondary effects such as the late-time integrated Sachs–Wolfe (ISW) effect~\cite{Hu:2004yd}. Nevertheless, CMB data remains crucial in background analyses because it helps break the degeneracies among cosmological parameters, such as $H_0$ and $\Omega_m$, and thus strengthen the overall constraints \cite{deside25}. In practice, for background-only analyses, the CMB can often be replaced by Gaussian priors on $(\theta_*, \omega_b, \omega_{bc})$; see \cite{deside25} for details. All background-only analyses presented in this paper adopt this simplification.  

\paragraph{Supernova constraints.}  
The tightest bounds on $(w_0, w_a)$ are provided by late-time probes, particularly type Ia supernovae and BAO. Type Ia SNe originate in binary systems where a white dwarf accretes mass from its companion until it approaches the Chandrasekhar limit of $1.44 M_\odot$, triggering a thermonuclear explosion of approximately standard luminosity ($M_V \sim -19.3$) \cite{Choudhuri_2010}. Their use as standard candles allows for precise reconstruction of the expansion history \cite{Riess_1998, Perlmutter_1999}. In practice, however, unambiguously classifying an event as a type Ia SN is observationally challenging \cite{Richards_2011}, leading to systematic differences among available SNe compilations \cite{efstathiou2025evolvingdarkenergysupernovae}.  

Currently, three major datasets of type Ia SNe; PantheonPlus \cite{Brout_2022}, Union3 \cite{rubin2025unionunitycosmology2000}, and DESY5 \cite{sánchez2024darkenergysurveysupernova};  are employed for cosmological inference in Ref.~\cite{deside25}. The corresponding posteriors are broadly consistent, but the statistical significance with which $\Lambda$CDM is excluded varies considerably across datasets. In particular, much of the constraining power of DESY5 and Union3 arises from very low-redshift measurements ($z < 0.1$) \cite{deside25}. Some studies have even argued that the DESI results could be explained as a local effect \cite{Gialamas_2025}, and have shown that excluding low-$z$ data removes the apparent tension, although the results remain tentative. To remain conservative while retaining some of the SNe constraining power, we focus exclusively on the PantheonPlus dataset, which yields the mildest deviation from $\Lambda$CDM \cite{deside25}. For computational efficiency, we use a compressed version of PantheonPlus, corresponding to a Gaussian likelihood of 10 binned measurements of the normalized Hubble parameter $H(z)/H_0$ spanning the redshift range $z \in [0.07, 2.5]$, as previously implemented in Ref.~\cite{Wolf_2024}.  

\paragraph{BAO constraints.}  
In the pre-recombination universe, baryons and photons were tightly coupled and behaved as a single oscillating fluid \cite{Baumann:2022mni}. These acoustic oscillations leave imprints in both the CMB and late-time clustering of matter. The baryonic oscillations source potential wells into which matter later collapses, producing a characteristic feature in the matter power spectrum. This imprint serves as a standard ruler, with the relevant scale set by the sound horizon at the drag epoch, $r_d \equiv r_s(z_d)$~\citep{Dodelson:2003ft,2005MNRAS.362..505C,SDSS:2005xqv,2011MNRAS.416.3017B}. 

DESI measures the BAO signal in both the transverse and radial directions, yielding the distance measures
\begin{align}
        D_M(z) = \int_0^z \frac{dz'}{H(z')}, 
    \qquad 
    D_H(z) = \frac{1}{H(z)},  \\
    D_V(z) = \big(z D_M^2(z) D_H(z)\big)^{1/3}.
\end{align}
In our analysis, we use DESI measurements of $D_V(z)/r_d$ and $D_M(z)/D_H(z)$, as quoted in Ref.~\cite{deside25}, across seven redshift bins spanning $z \in [0.295, 2.33]$, obtained by combining multiple tracers of large-scale structure, including galaxies and quasars. As with the SNe datasets, deviations from $\Lambda$CDM are driven by late-time measurements. It has been demonstrated that if the low-$z$ measurements ($z < 0.8$) of DESI are replaced by SDSS measurements, consistency with $\Lambda$CDM is fully restored \cite{chudaykin2024modifiedgravityinterpretationevolving}. 

\paragraph{Combined constraints.}  
In Fig.~\ref{fig:Constraints1}, we illustrate the constraints on $(w_0, w_a)$ derived from the PantheonPlus SNe, \textit{Planck} 2018 CMB~\citep{Planck:2018vyg}, and DESI BAO datasets, both individually and in combination. While the CMB alone does not provide significant bounds on dark-energy parameters, it plays an essential role in breaking degeneracies in the background cosmology and thereby sharpens the joint constraints. The full set of background probes measure $w_0$ and $w_a$ to be
\begin{equation}
w_0 = -0.84 \pm 0.06, \qquad w_a = -0.62 \pm 0.21   
\end{equation}
which signifies a $2.8\sigma$ deviation from $\Lambda$CDM~\cite{deside25}.

For this figure, we analyzed the different dataset combinations by performing separate MCMC runs for each case. In subsequent sections, however, we approximate all posteriors as Gaussian and combine cosmological probes by adding the corresponding Fisher matrices. The full MCMC analyses allow us to extract best-fit Gaussian approximations to the background-only posteriors, which are then used consistently throughout the rest of this work. 

\subsection{The Matter Power Spectrum}
The background measurements discussed thus far already place meaningful constraints on the allowed $(w_0, w_a)$ parameter space, with current observations indicating potential deviations from $\Lambda$CDM. To further tighten these constraints and probe the microphysical properties of dark energy, we extend our analysis to include detailed measurements of the matter power spectrum. In particular, we focus on the additional information provided by the kinetic Sunyaev–Zel’dovich (kSZ) effect, obtained by combining CMB observations with overlapping galaxy surveys. The kSZ effect and the survey specifications underlying our forecasts are described in detail in the following section.

\begin{figure}
    \centering
    \includegraphics[width=1.0\linewidth]{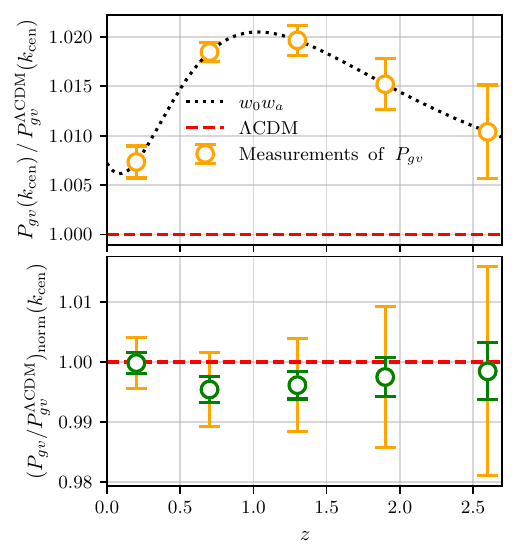}
    \caption{ Forecasted measurements of the galaxy–velocity cross-spectrum $P_{gv}$ based on LSST and CMB-S4 specifications, with error estimates from \cite{smith2018ksztomographybispectrum}. \textbf{Top:} scale-independent modifications to the spectrum, illustrated using small-scale modes. \textbf{Bottom:} scale-dependent modifications, shown for large-scale modes, normalized on small scales. In green we also show how the SNR increases if the CMB survey parameters $(s_w, \theta_{\rm FWHM})$ are reduced by a factor of two relative to those described in Eq.~\eqref{eq:CMBspec}. }
    \label{fig:PowerSpectra}
\end{figure}

To illustrate the potential of such measurements, we consider the galaxy–velocity cross-spectrum $P_{gv}$, whose errors are derived after marginalizing over the galaxy–galaxy ($P_{gg}$) and velocity–velocity ($P_{vv}$) spectra~\cite{smith2018ksztomographybispectrum}. In Fig.~\ref{fig:PowerSpectra}, we show fiducial predictions for $P_{gv}$ based on the DESI best-fit values of $(w_0, w_a)$, compared against the $\Lambda$CDM case. We highlight two distinct effects: The scale-independent effect, quantified by $P_{gv}$ averaged over small-scale modes in the bin $k \in [10^{-2}, 10^{-1}]\,\mathrm{Mpc}^{-1}$ and the scale-dependent effect, evaluated from the large-scale bin $k \in [k_{\min}, 2 \times 10^{-3}]\,\mathrm{Mpc}^{-1}$ and normalized on small scales, where $k_{\min} = \pi / V^{1/3}$ is determined by the survey volume $V$~\cite{smith2018ksztomographybispectrum}.  

As emphasized above, the scale-independent effect is considerably stronger and in principle appears readily accessible to observations. In practice, however, $P_{gv}$ is not directly observable, but is instead obscured by galaxy bias and other observational systematics that must be marginalized over. Redshift-space distortions (proportional to $f\mu^2$) partially break this degeneracy, allowing the overall amplitude of the power spectrum to constrain cosmological parameters.  

By contrast, the scale-dependent effect is intrinsically small, lying well below the sensitivity of upcoming surveys. These large-scale modes do contain non-trivial redshift information that can improve cosmological constraints, but neither current survey volumes nor measurement precision are sufficient to fully access the horizon-scale signatures most directly connected to the microphysics of dark energy. We also demonstrate qualitatively how these errors could be improved by considering futuristic CMB survey specifications in Fig.~\ref{fig:PowerSpectra}. Moreover, on these scales measurement uncertainties are fundamentally dominated by cosmic variance, which poses a severe limitation.  

This is precisely where the kSZ effect becomes especially powerful: by enabling sample variance cancellation, kSZ tomography provides a way to eliminate cosmic variance in velocity field measurements \cite{M_nchmeyer_2019}. In doing so, it opens up an otherwise inaccessible observational window on the scale-dependent effects of dark-energy perturbations.  

\section{Forecasts}\label{secIV}

\subsection{Experiments}

To forecast the constraints attainable from future measurements of the kSZ effect, we adopt survey specifications consistent with those expected from the upcoming CMB-S4 and LSST experiments.  

\paragraph{CMB experiment.}  
Our baseline CMB survey is modeled on the planned CMB-S4 experiment. We describe the beam-deconvolved noise spectrum of the foreground-cleaned CMB map, $N(\ell)$, as Ref.~\cite{M_nchmeyer_2019}
\begin{equation}
    N(\ell) = s_w^2 \exp \!\left[ \frac{\ell(\ell + 1)}{8 \ln 2}\,\theta_{\rm FWHM}^2 \right],
\end{equation}
where $s_w$ denotes the effective white-noise level and $\theta_{\rm FWHM}$ the beam full width at half maximum (FWHM). Following \cite{M_nchmeyer_2019}, we adopt fiducial values
\begin{equation}
    s_w = 1 \, \mu\mathrm{K}\,\mathrm{arcmin}, 
    \qquad 
    \theta_{\rm FWHM} = 1.5 \,\mathrm{arcmin}.
    \label{eq:CMBspec}
\end{equation}

\paragraph{Galaxy survey.}  
For the large-scale structure component, we adopt specifications consistent with the Large Synoptic Survey Telescope (LSST). We use forecasted values based on projected survey parameters \cite{M_nchmeyer_2019}, which are summarized in Table~\ref{tab:LSST}. The photometric redshift error is modeled as
\begin{equation}
    \sigma_z = 0.03\,(1+z),
\end{equation}
in line with previous studies \cite{M_nchmeyer_2019}.  

The survey is divided into five redshift bins, allowing the redshift evolution of the power spectrum to be captured in our inference. Throughout, we consider Fourier modes in the range
\begin{equation}
    k \in \big[\pi/V^{1/3}, \, 0.1\,\mathrm{Mpc}^{-1}\big],
\end{equation}
where $V$ denotes the comoving volume of the corresponding redshift bin. The momentum cutoff $k_{\min} = 0.1 \,\mathrm{Mpc}^{-1}$ ensures that we remain well within the linear regime. 

\begin{table}[h]
    \centering
    \begin{tabular}{cccccc}
        \hline
        bin & $z_{\text{min}}$ & $z_{\text{max}}$ & galaxy bias $b_g$ & galaxy density $n_g$ & volume $V$ \\
        \hline
        1 & 0.0 & 0.4 & 1.05 & 0.05 $\,\,\mathrm{Mpc}^{-3}$ & 5.2 $\,\,\mathrm{Gpc}^3$ \\
        2 & 0.4 & 1.0 & 1.37 & 0.02 $\,\,\mathrm{Mpc}^{-3}$ & 43.6 $\,\,\mathrm{Gpc}^3$ \\
        3 & 1.0 & 1.6 & 1.79 & 0.006 $\,\,\mathrm{Mpc}^{-3}$ & 75.9 $\,\,\mathrm{Gpc}^3$ \\
        4 & 1.6 & 2.2 & 2.22 & 0.0015 $\,\,\mathrm{Mpc}^{-3}$ & 89.3 $\,\,\mathrm{Gpc}^3$ \\
        5 & 2.2 & 3.0 & 2.74 & 0.0003 $\,\,\mathrm{Mpc}^{-3}$ & 119.9 $\,\,\mathrm{Gpc}^3$ \\
        \hline
    \end{tabular}
    \caption{Fiducial values of galaxy bias, number density, and comoving volume for five redshift bins, consistent with projected LSST specifications \cite{M_nchmeyer_2019}.}
    \label{tab:LSST}
\end{table}

\subsection{The kinetic Sunyaev–Zel’dovich Effect}

The anisotropy in the CMB maps induced by the kSZ effect along a line of sight $\boldsymbol n$ is given by
\begin{equation}
\frac{\delta T(\boldsymbol{\theta})}{T_{\rm CMB}}\Big|_{\rm kSZ} 
    = K(z_\star) \int_0^L {\mathrm d}r \, q_\parallel(\boldsymbol{x}),
\end{equation}
where $T_{\rm CMB}$ is the mean CMB temperature, $\boldsymbol{x}\equiv\chi_\star\boldsymbol{\theta}+r\hat{\boldsymbol{r}}$ is the conformal distance to a large-scale structure shell at redshift $z=z_\star$, $\boldsymbol{\theta}$ is the angular direction on the sky, $\hat{\boldsymbol{r}}$ is the unit vector in the radial direction and $q_\parallel(\boldsymbol{x})=\delta_e(\boldsymbol{x})v_\parallel(\boldsymbol{x})$ is the electron-momentum field, projected onto the radial direction. The kSZ radial weight function satisfies
\begin{equation}
    K(z)=-\sigma_T n_{e,0} x_e(z)e^{-\tau(z)}(1+z)^2
\end{equation}
where $\sigma_T$ the Thomson cross-section, $n_e$ the free-electron density, $x_e(z)$ the ionization fraction, and $\tau(z)$ is the optical depth at redshift $z$. 

In this work we exploit the cosmological information encoded in kSZ tomography, combining CMB measurements with a tracer of the electron distribution (such as galaxies) to reconstruct the remote velocity field on large scales. This velocity field contains information on the clustering properties of dark energy: when $w \neq -1$, the growth rate of structure differs from its $\Lambda$CDM prediction and acquires a scale dependence, leaving characteristic signatures in velocity correlations. Crucially, the remote velocity field is an unbiased tracer of the total matter density. Its cross-correlation with biased galaxy density fields enables the use of sample-variance cancellation ~\cite{M_nchmeyer_2019,Hotinli:2019wdp,AnilKumar:2022flx,Kumar:2022bly}, allowing scale-dependent deviations from $\Lambda$CDM to be extracted at high precision on large scales.  

Following \cite{smith2018ksztomographybispectrum}, the reconstruction noise for a large-scale velocity mode of wavenumber $k_L$ at comoving distance $\chi_*$ is
\begin{equation}
N^{vv}_{\rm rec}(k_L,\mu) 
= \frac{2\pi \chi_*^2}{\mu^2 K_*^2}
\left[ \int \mathrm d k_s \, k_s \, 
\frac{P_{ge}(k_s)^2}{P^{\rm obs}_{gg}(k_s)\, C_{\ell=k_s \chi_*}^{TT, \rm obs}} \right]^{-1},
\end{equation}
where the parameter $\mu = \hat k \cdot \hat n$ denotes the cosine of the angle between the mode and the line of sight.  

In Fig.~\ref{fig:PowerSpectra}, we illustrate the constraining power of future kSZ measurements by comparing the reconstructed galaxy–velocity power spectrum $P_{gv}$ (with errors determined as above) in two fiducial cosmologies. The first corresponds to $\Lambda$CDM, $(w_0, w_a) = (-1,0)$, while the second adopts $(w_0, w_a)$ values equal to the best-fit combination of BAO, CMB, and SNe data. All other cosmological parameters are held fixed. This figure demonstrates that kSZ tomography provides a powerful means of constraining dark energy, once background parameters are marginalized over, by probing perturbation-level signatures that are inaccessible to background-only analyses.

\begin{figure}
    \centering
    \includegraphics[width=1\linewidth]{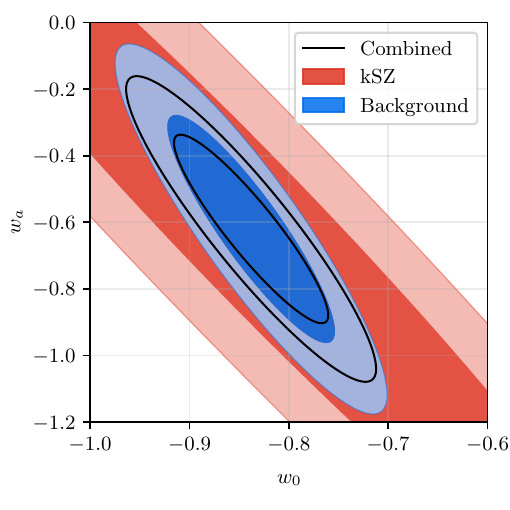}
    \caption{ Constraints on $(w_0, w_a)$ from a joint Fisher analysis of background (BAO, CMB and SNe) and kSZ ($P_{gg}$, $P_{gv}$ and $P_{vv}$) probes. The background posterior is taken from the Gaussian approximation to the MCMC posteriors derived in previous sections. Background and kSZ constraints are combined by adding the corresponding Fisher matrices. We show the full Fisher matrix, as well as constraints derived using more futuristic CMB specifications, in the appendix. }
    \label{fig:ResultkSZ}
\end{figure}

\subsection{Fisher Matrix}

We consider a joint measurement of small-scale CMB anisotropies and galaxy number counts over the same region of sky. As shown in Ref.~\cite{smith2018ksztomographybispectrum}, the cosmological information contained in such a survey can be described in terms of a data vector consisting of velocity modes $v(\boldsymbol k)$, reconstructed via the kSZ effect, and galaxy overdensity modes $\delta^g(\boldsymbol k)$. Both are biased tracers of the underlying matter perturbations $\delta(\boldsymbol k)$. In linear theory, these relations are
\begin{equation}
    \label{eq:Reconstruction}
    \delta^g(\boldsymbol k) = (b_g + f\mu^2)\,\delta(\boldsymbol k), 
    \qquad 
    v(\boldsymbol k) = b_v \frac{f a H}{k}\, \delta(\boldsymbol k),
\end{equation}
where $b_g$ is the galaxy bias, $f = \partial \ln \delta / \partial \ln a$ the growth rate, and $b_v$ the velocity reconstruction bias arising from the kSZ optical-depth degeneracy \cite{smith2018ksztomographybispectrum}. The $f \mu^2$ term represents the Kaiser redshift-space distortion (RSD) contribution, where $\mu = k_r/k$ is the cosine of the angle between the mode and the line of sight. While this expression accounts only for the leading-order term in the bias expansion, it suffices for our analysis, as we consider only linear scales.

Following \cite{M_nchmeyer_2019}, the signal and noise covariance matrix of the survey takes the form
\begin{equation}
    \mathbf C(k,\mu,z) =
    \begin{pmatrix}
        P_{vv} + N_{vv}^{\rm rec} & P_{vg} \\
        P_{vg} & P_{gg} + N_{gg}
    \end{pmatrix},
\end{equation}
where, following specifications in  \cite{lsstsciencecollaboration2009lsstsciencebookversion}, the galaxy noise is modeled as shot noise with photometric redshift smearing,  
\begin{equation}
    N_{gg} = \frac{1}{W \cdot n_{\rm gal}}, 
    \qquad 
    W = \exp\!\left[-(0.03\,\mu)^2 \left(\tfrac{k}{aH}\right)^2\right].
\end{equation}

We compute the small-scale galaxy–electron and galaxy–galaxy power spectra, $P_{ge}$ and $P_{gg}$, using the publicly available code $\mathtt{hmvec}$ \cite{2025ascl.soft02030M}, which incorporates one- and two-halo contributions. To capture uncertainties in galaxy clustering, we introduce 10 nuisance parameters corresponding to galaxy biases $b_g$ and velocity biases $b_v$ across redshift bins. These nuisance parameters are marginalized over when deriving cosmological constraints.  

The Fisher matrix for the cosmological parameters $(w_0, w_a)$ is then
\begin{equation}
    F_{ab} = \frac{V}{2} \int_{k_{\min}}^{k_{\max}} \int_{-1}^1 
    \frac{k^2\, \mathrm d k\, \mathrm d \mu}{(2 \pi)^2}\,
    \operatorname{Tr}\!\left[
        \frac{\partial \mathbf C_*}{\partial \theta_a}\,
        \mathbf C_*^{-1}\,
        \frac{\partial \mathbf C_*}{\partial \theta_b}\,
        \mathbf C_*^{-1}
    \right],
\end{equation}
see Refs.~\cite{smith2018ksztomographybispectrum, M_nchmeyer_2019} for details. This is the standard Fisher matrix for a Gaussian likelihood, with the integrals implementing a sum over all Fourier modes accessible within the survey volume. Each redshift bin contributes its own Fisher matrix, including two additional nuisance parameters ($b_g$, $b_v$), with fiducial values of $b_g$ taken from Table~\ref{tab:LSST} and $b_v = 1$ fiducially. The Fisher matrices from individual bins are then combined, treating the bins as statistically independent.

\section{Results}\label{secV}

The constraints on $(w_0, w_a)$ obtained from the Fisher matrix analysis described in the previous section are shown in Fig.~\ref{fig:ResultkSZ}. Including kSZ information leads to an improvement in the constraints, though only marginally. { Adding velocity information ($P_{gv}$ and $P_{vv}$) to galaxy clustering ($P_{gg}$) alone improves constraints on background parameters such as $w_0$ and $w_a$ by only $5$ to $10\%$. This is because constraints on these parameters are dominated by scale-independent measurements of the growth rate from a full-shape analysis, which are only marginally improved by including velocity spectra. The more substantial gains from kSZ arise for scale-dependent modulations of the power spectrum, as explored in Sec.~\ref{secV}.}

Importantly, however, the degeneracy direction of the kSZ constraints differs slightly from that of the background probes. This distinction arises because expansion history measurements, such as those of $H(z)$, are sensitive only to the integrated equation of state,
\begin{equation}
    \Omega_{\rm de}(a) = \Omega_{\rm de,0} \exp \left[-3 \int (1 + w)\, \mathrm d \ln a \right],
\end{equation}
whereas perturbation-based observables are additionally sensitive to the averaged equation of state \cite{PhysRevD.59.063005},
\begin{equation}
    \overline w = \frac{\int \Omega_{\rm de}(a)\,w(a)\,\mathrm d a}{\int \Omega_{\rm de}(a)\,\mathrm d a}.
\end{equation}
The dependence of $\overline w$ on $w(a)$ differs from that of $\Omega_{\rm de}(a)$, and this naturally leads to different degeneracy directions. Crucially, this effect is expected irrespective of the specific phenomenological parameterization used for dark energy.  

\begin{figure}
    \centering
    \includegraphics[width=1\linewidth]{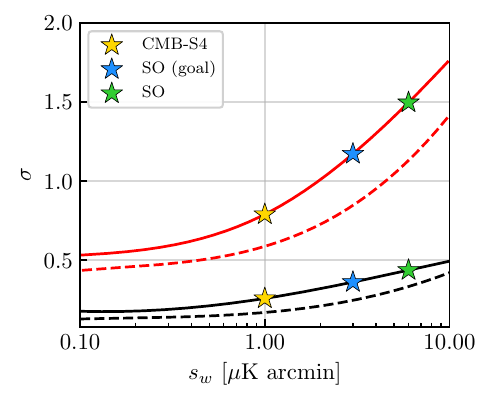}
    \caption{Uncertainties in the dark-energy equation-of-state parameters $(w_0, w_a)$ as a function of CMB survey specifications. The red lines show constraints on $w_a$, while the black line shows $\sigma_{w_0}$. Here, the dashed line assumes the very futuristic configuration $\theta_{\rm FWHM} \simeq 0.2'$, while the solid line sets $\theta_{\rm FWHM} \simeq 1.4'$. As expected, reducing instrumental noise improves the constraints, though the dependence on the white-noise level $s_w$ saturates near $s_w \simeq 0.3 \, \mu \mathrm {K}\, \mathrm{arcmin}$. { This also illustrates how observations from the Simons Observatory, with $\theta_{\rm FWHM} \simeq 1.4'$ and $s_w \simeq 6\, \mu \mathrm{K}\, \mathrm{arcmin}$ \cite{Ade:2018sbj}, can nonetheless place meaningful constraints on $w_0$, while $w_a$ remains largely unconstrained at this sensitivity.}}
    \label{fig:Improvements}
\end{figure}

The constraints shown in Fig.~\ref{fig:ResultkSZ} can be substantially improved by adopting more ambitious survey specifications. As demonstrated in Fig.~\ref{fig:Improvements}, reducing the beam full-width at half-maximum (FWHM) $\theta_{\rm FWHM}$ and lowering the effective white-noise level $s_w$ markedly enhances the constraining power of kSZ measurements, elevating them to a level comparable with traditional background probes. In this regime, incorporating kSZ data into cosmological inference delivers meaningful gains over current limits.

{While kSZ measurements are not yet competitive with traditional background probes, they provide a complementary probe of dark energy through the velocity field, accessing perturbation-level signatures that geometric measurements do not directly capture.} This makes them {especially useful} as a consistency test: the joint combination of kSZ + CMB + BAO yields noticeably tighter constraints than those obtained from CMB and BAO alone. While this combination does not yet match the constraining power of the benchmark set SNe + CMB + BAO, it provides a complementary avenue that can help mitigate systematic uncertainties and interpretive challenges associated with supernova-based measurements. Moreover, if the $(w_0,w_a)$ parameterization is to be regarded as a proxy for the microphysical properties of dark energy, then any viable model must reproduce not only the background expansion history, but also the observed perturbations of matter. In this sense, kSZ observations offer a { direct} means of testing dark-energy models beyond the background level and provide a pathway to probing the underlying microphysics directly. 

To explore this possibility, we investigate how kSZ measurements can constrain dark energy using the shape of the matter power spectrum. Inspired by Ref.~\cite{Ma_1999} and Eq.~\eqref{eq:Decay}, we model the effect of dark energy by parameterizing the density fluctuations as
\begin{equation}
    \begin{cases}
        \delta(\boldsymbol k) = \Sigma(\boldsymbol k)\,\delta^{\Lambda{\rm CDM}}(\boldsymbol k), \\[6pt]
        \displaystyle \Sigma(\boldsymbol k) = \Sigma_0 \frac{\alpha + q^2}{1 + q^2}, 
        \qquad q = \frac{\ell_s k}{aH}
    \end{cases}.
    \label{eq:fit}
\end{equation}
Here $\delta^{\Lambda{\rm CDM}}(\boldsymbol k)$ are the matter fluctuations in a $(w_0,w_a)=(-1,0)$ universe, with all other cosmological parameters fixed. The parameter $\Sigma_0$ quantifies the scale-independent modification of the power spectrum, which, as described in Sec.~\ref{secII}, arises entirely due to background effects. Thus, background measurements can be used to impose a Gaussian prior on $\Sigma_0$,
\begin{equation}
    \Sigma_0 \sim \mathcal N(1.0092, (0.0112)^2),
\end{equation}
thereby reducing degeneracies between $\Sigma_0$ and $\alpha$.\footnote{Specifically, we compute $\Sigma_0$ by comparing the small-scale $\Lambda$CDM power spectrum with that of the $w_0w_a$ model across all $(w_0,w_a)$ values sampled in our previous MCMC analysis. We then fit a Gaussian to the resulting distribution of $\Sigma_0$.} We note that in general $\Sigma_0$ and $\alpha$ can have highly non-trivial redshift dependence. To simplify the analysis we therefore focus on the constraints that can be derived from considering a single redshift bin. Here, we forecast results based on the baseline 2  specifications discussed in Ref.~\cite{M_nchmeyer_2019}. 

The parameters $(\alpha,\ell_s)$ fully encapsulate deviations from $\Lambda$CDM induced by dark-energy microphysics. In the $k \to 0$ limit, $\Sigma/\Sigma_0 \to \alpha$, so $\alpha$ measures the overall amplitude of deviations from $\Lambda$CDM on ultra-large scales. For $q \gg 1$ one instead finds $\Sigma/\Sigma_0 \rightarrow 1$, whereby we impose that dark energy perturbations are negligible on small scales. For adiabatic initial conditions, one expects 
\begin{equation}
    \alpha - 1 \sim (1+ \overline w), 
\end{equation}
which explains the suppression of growth on large scales observed for the best-fit $(w_0,w_a)$ values in Fig.~\ref{fig:PowerSpectra}. A direct measurement of $\alpha$ would also help bridge the gap between $(w_0,w_a)$ constraints from observations of the expansion history and fundamental physics: only in models that allow a phantom crossing can $\alpha < 1$, making such a detection a sharp discriminator between scenarios.  

\begin{figure}
    \centering
    \includegraphics[width=1\linewidth]{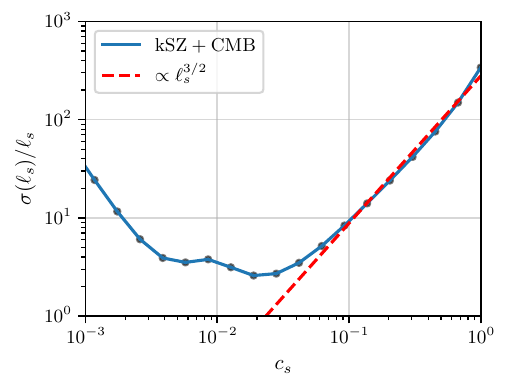}
    \caption{Relative error on the sound horizon as a function of the fiducial sound speed, computed from a Fisher analysis of a galaxy survey $\delta^g$ and reconstructed velocity modes $v$, as well as a CMB prior on all background parameters.  Lower sound speeds shift the sound horizon to smaller scales, where it is more easily measured, thereby enhancing observational significance. We demonstrate how the scaling seen above can be motivated analytically in the appendix. This improvement saturates, and reverses, once the sound horizon falls below the smallest accessible scale $\simeq 1/k_{\max}$.}
    \label{fig:SoundSpeed}
\end{figure}

\begin{figure}
    \centering
    \includegraphics[width=1\linewidth]{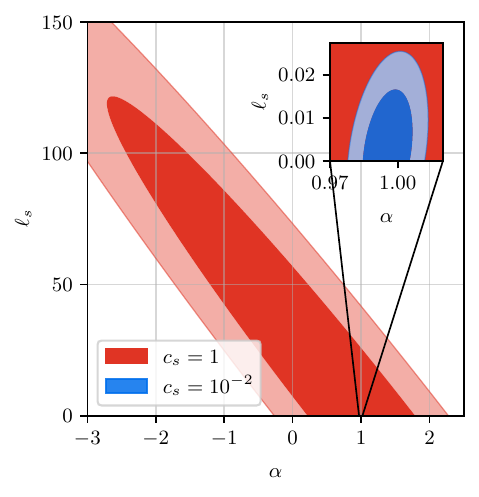}
    \caption{Constraints on the phenomenological parameters $(\alpha, \ell_s)$ from kSZ measurements. For the canonical sound speed $c_s = 1$, current survey sensitivities do not yield meaningful constraints, since the most relevant (super-horizon) modes lie on scales inaccessible to existing observations. However, for smaller sound speeds, where the sound horizon shifts to sub-horizon scales, kSZ measurements can provide significant and informative constraints. }
    \label{fig:SoundSpeed}
\end{figure}

The parameter $\ell_s$ sets the characteristic scale (relative to the Hubble horizon) at which dark-energy clustering begins to significantly modify the matter power spectrum. In a fluid description, the size of the dark-energy sound horizon (relative to the Hubble horizon) can be written as \cite{Creminelli_2009}
\begin{equation}
    \ell = aH \int c_s \,\frac{dt}{a},
\end{equation}
where $c_s$ denotes the dark-energy sound speed. This expression does not coincide exactly with the role of $\ell_s$ in our fitting relation in Eq.~\eqref{eq:fit}, where $\ell_s$ instead specifies the scale at which dark-energy perturbations have decayed to half their initial amplitude. Numerically, we find that these two scales differ by a nearly redshift-independent factor $\ell/\ell_s \simeq 9.4$, with only a weak dependence on the averaged equation of state $\overline w$.

We do not impose priors on $(\alpha,\ell_s)$,\footnote{Beyond the physical requirement of $\ell_s \ge 0$. } as our aim is to assess how well kSZ measurements alone can constrain these parameters in a model-agnostic way, independent of the chosen background parameterization. In this sense, consistency between the observed values of $(\alpha,\ell_s)$ and those predicted by, for example, a quintessence model constrained by the background data would provide an important validation of dark-energy physics. The forecasted constraining power for $(\alpha,\ell_s)$, based on the kSZ survey specifications described in Ref.~\cite{M_nchmeyer_2019}, is shown in Fig.~\ref{fig:SoundSpeed}. We note that these are based on the fiducial choice \begin{equation}(\alpha, \ell_s) = (0.995, 0.237),\end{equation} which corresponds to the best-fit parameters given the DESI best-fit values for $(w_0, w_a)$. 

Choosing $\ell_s = 0.237$ corresponds to the canonical sound speed $c_s = 1$, in which case the sound horizon lies on horizon scales and is therefore difficult to probe observationally. Sensitivity improves substantially if the sound horizon shifts to scales smaller than the Hubble radius. To illustrate this, we compute the relative error on $\ell_s$ for a range of fiducial $c_s$ values (see Fig.~\ref{fig:SoundSpeed}). For illustrative purposes, we keep $\alpha$ fixed in this calculation, although in general $\alpha$ and $c_s$ are not independent; their relation depends on the specific underlying dark energy model. It is worth noting that small or even vanishing sound speeds are both theoretically consistent and, in fact, required by effective field theory considerations if a phantom crossing occurs \cite{Creminelli_2009}. The prospect of detecting dark energy in the small $c_s$ regime is particularly intriguing and deserves further investigation.

\section{Discussion}\label{secVI}

The accelerating expansion of the Universe remains one of the most compelling and least understood phenomena in modern cosmology. While background observables---SNe, BAO, and the CMB---have established the existence of dark energy and continue to refine the $(w_0, w_a)$ constraints, they probe only the smooth expansion history. A complete characterization of dark energy must also include its \emph{perturbations}, which influence structure formation and leave subtle, scale-dependent signatures in the matter power spectrum. Motivated by this, we have explored how the combination of galaxy clustering and kSZ velocity reconstruction can be used to access these perturbation-level effects \emph{and} to improve constraints on the background parameters of dark energy. The kSZ effect, produced by the scattering of CMB photons off moving free electrons, enables reconstruction of the large-scale radial velocity field---a direct tracer of the total matter distribution. By combining this reconstructed velocity field with galaxy density maps, one can measure the joint spectra $(P_{gg}, P_{gv}, P_{vv})$ and exploit sample-variance cancellation to isolate departures from $\Lambda$CDM growth. Using this framework, we forecast the joint constraints attainable with forthcoming LSST and CMB-S4-like surveys, quantifying how galaxy and velocity spectra complement background probes and assessing whether dark-energy perturbations---encoded in parameters such as the sound speed $c_s$ or the scale $\ell_s$---might become observationally accessible.

Our forecasts show that incorporating kSZ tomography alongside galaxy clustering provides a complementary and empirically valuable avenue for testing dark energy. While kSZ information alone does not yet yield a dramatic improvement over background probes, it introduces qualitatively new sensitivity to the growth of structure and to the gravitational effects of dark-energy perturbations. The kSZ power spectra $(P_{gg}, P_{gv}, P_{vv})$ probe the time- and scale-dependence of the velocity field and its cross-correlation with galaxy density, constraining both the background expansion rate and the rate of structure growth.

Including kSZ data tightens constraints on the background equation-of-state parameters $(w_0, w_a)$ while rotating their degeneracy directions relative to those of purely geometric probes. The combination of CMB, BAO, and SNe data with kSZ-derived velocity information reduces the marginalized uncertainties on $(w_0, w_a)$ by tens of percent (depending on survey specifications) and provides a valuable cross-check. These results emphasize that even moderate-resolution kSZ measurements can contribute meaningful information to background inference.

A second outcome concerns the detectability of dark-energy perturbations. We introduced a simple two-parameter model capturing their leading-order impact on the matter power spectrum,
\[
\delta(k)=\Sigma(k)\,\delta^{\Lambda\mathrm{CDM}}(k), \quad 
\Sigma(k)=\Sigma_0\,\frac{\alpha+q^2}{1+q^2}, \quad q=\frac{\ell_s k}{aH},
\]
which separates a scale-independent background component $\Sigma_0$ from a scale-dependent modulation governed by $(\alpha,\ell_s)$. In this framework, the canonical quintessence limit ($c_s=1$) places the sound horizon $\ell_s$ near the Hubble scale, producing sub-percent modifications to the power spectrum on horizon scales. Consequently, such perturbation-level signatures remain observationally elusive for realistic survey volumes and noise levels, becoming significant only in models with reduced sound speeds ($c_s\!\ll\!1$) where the relevant turnover shifts to sub-horizon scales. { A caveat to this conclusion arises in models where dark energy and dark matter are coupled through momentum transfer \cite{Pourtsidou:2013nha, Xu:2013jma}. In such models, the interaction acts as a drag force in the Euler equation, modifying the velocity field at first order while affecting the density field only indirectly through the continuity equation. As a result, peculiar velocity measurements---and by extension kSZ tomography---could provide stronger constraints on dark energy--dark matter coupling than density-based probes alone \cite{Xu:2013jma}. We leave the investigation of this effect to future work.}

We restricted our analysis to linear scales ($k \lesssim 0.1~{\rm Mpc}^{-1}$), where perturbation theory remains accurate, and marginalized over a minimal set of nuisance parameters describing galaxy bias $b_g$ and velocity-reconstruction bias $b_v$. We did not attempt to marginalize over model parameters concerning halo-galaxy connections, baryonic feedback, or uncertainties in the galaxy–electron cross-spectrum (except for the optical-depth bias introduced above), whose accurate treatment would require detailed hydrodynamical simulations. The linear sound-speed prescription captures only the leading behavior of dark-energy perturbations and cannot describe clustering away from the fluid limit. In practice, survey masks and other systematics can also suppress sensitivity to the largest modes where dark-energy perturbations are expected to appear. The forecasts presented here should therefore be interpreted as upper limits on the constraining power of near-term surveys. For canonical sound speed ($c_s = 1$), dark-energy perturbations induce only percent-level modulations on horizon scales; even under optimistic assumptions about instrumental noise and beam size, such signals will remain difficult to isolate. Hence, while kSZ tomography is conceptually well-suited to probing perturbation-level physics, its near-term value lies primarily in validating $\Lambda$CDM and cross-checking background inferences rather than delivering decisive detections of dark-energy microphysics.

Taken together, these results suggest that kSZ tomography represents a realistic but marginal step toward a more complete empirical characterization of dark energy. Its main strength is complementarity: kSZ observables probe structure growth in a way that is not degenerate with the background expansion history. The resulting constraints in the $(w_0,w_a)$ plane have orientations distinct from those of purely geometric probes, so the addition of kSZ probes provides an important internal consistency test. Even if the current signal-to-noise remains modest, such cross-validation may become increasingly valuable as background measurements reach sub-percent precision and systematic control becomes the dominant challenge.

Several avenues could meaningfully extend the reach of this framework. Lowering CMB white noise ($s_w \lesssim 0.3~\mu{\rm K\,arcmin}$) and achieving finer beam resolution ($\theta_{\rm FWHM} \lesssim 1'$) would enhance velocity-reconstruction fidelity and make kSZ contributions more complementary with some background probes. Expanding survey area and redshift coverage, together with finer tomographic binning, would increase the number of large-scale modes and reduce the cosmic-variance ceiling that presently limits sensitivity to dark-energy perturbations. Joint analyses incorporating weak-lensing shear or convergence fields can further suppress sample variance and strengthen cross-checks between velocity- and density-based growth measurements. Improved modeling of the galaxy–electron power spectrum $P_{ge}$ and independent calibration of the velocity-bias parameter $b_v$---for instance through FRB-based optical-depth measurements---will be essential to fully exploit the statistical potential of future data. Finally, moving beyond the specific $(\alpha,\ell_s)$ template, a more general phenomenological description of large-scale departures from $\Lambda$CDM---expressed in terms of effective clustering amplitudes, scale-dependent growth responses, or EFT-inspired operators---may offer a broader and more flexible framework for testing the microphysics of cosmic acceleration, as has already been proposed in \cite{Gubitosi_2013, Bellini_2014, Creminelli_2010}. 

While the path toward detecting dark-energy perturbations remains steep, the combination of improved CMB data, expanded galaxy surveys, and continued methodological refinement keeps the prospect of unveiling dark-energy microphysics alive. The analysis presented here outlines both the practical challenges and the conceptual opportunities in transforming kSZ tomography from a consistency check into a precision probe of cosmic acceleration.

\section*{Acknowledgments }
We thank Wayne Hu and Mathew Madhavacheril for useful discussions. JA would like to thank the generous support of the Perimeter Scholars International scholarship. Research at Perimeter Institute is supported in part by the Government of Canada through the Department of Innovation, Science and Economic Development and by the Province of Ontario through the Ministry of Colleges and Universities.

\bibliography{de_pert_ksz}

\appendix
\section{Analytic Fisher}
The Fisher matrix in a given redshift bin $z$ is 
\begin{equation}
    F_{ab} = \frac{V}{2} 
    \int_{k_{\min}}^{k_{\max}} \int_{-1}^1 
    \frac{k^2\, \mathrm{d}k\, \mathrm{d}\mu}{(2\pi)^2}\,
    \mathrm{Tr}\!\left[
        \frac{\partial \mathbf{C}}{\partial \theta_a}\,
        \mathbf{C}^{-1}\,
        \frac{\partial \mathbf{C}}{\partial \theta_b}\,
        \mathbf{C}^{-1}
    \right],
\end{equation}
where the covariance matrix is given by
\begin{equation}
\mathbf{C}(k,\mu,z) =
    \begin{pmatrix}
        P_{vv} + N_{vv}^{\rm rec} & P_{vg} \\
        P_{vg} & P_{gg} + N_{gg}
    \end{pmatrix}.
\end{equation}
The various power spectra entering this matrix are modeled as
described in Eq.~(\ref{eq:Reconstruction})
and we model the effects of dark energy analytically via
\begin{equation}
    P_{mm} = \Sigma(k)^2\,P_{mm}^{\Lambda{\rm CDM}}, \qquad 
    \Sigma(k) = \Sigma_0\,\frac{\alpha + q^2}{1 + q^2}, 
\end{equation}
where as before $q = {\ell_s k}/{aH}.$

\subsection{Per-mode Fisher information}
In the idealized limit where $N_{vv}^{\rm rec} \!\approx\! 0$ and $N_{gg} \!\approx\! 0$, one finds $\mathbf{C} = \Sigma^2(k)\mathbf{C}_0$, where $\mathbf{C}_0$ is independent of $(\ell_s,\alpha)$.  
The Fisher matrix per mode then becomes
\begin{equation}
    F_{ab}(k) = 4V\,
    \frac{\partial}{\partial \theta_a}\!\log\Sigma(k)\,
    \frac{\partial}{\partial \theta_b}\!\log\Sigma(k),
\end{equation}
where we used $\mathrm{Tr}(\mathbf{1}) = 2$ (and treat $\Sigma_0$ as fixed at its fiducial value).
The relevant derivatives are
\begin{equation}
    \frac{\partial}{\partial \alpha} \log\Sigma 
    = \frac{1}{\alpha + q^2},
\end{equation}
\begin{equation}
    \frac{\partial}{\partial \ell_s} \log\Sigma 
    = \frac{2q^2(1-\alpha)}{\ell_s(1+q^2)(\alpha + q^2)}.
\end{equation}
This yields
\begin{align}
    F_{\alpha\alpha}(k) &= \frac{4V}{(\alpha + q^2)^2}, \\
    F_{\ell_s\ell_s}(k) &= 4V\!\left[\frac{2q^2(1-\alpha)}{\ell_s(1+q^2)(\alpha + q^2)}\right]^{\!2}, \\
    F_{\alpha\ell_s}(k) &= \frac{8Vq^2(1-\alpha)}{\ell_s(1+q^2)(\alpha + q^2)^2}.
\end{align}
We note the scaling behavior
$
    F_{ij} \propto q^{-4}.
$
Including the $k^2$ weighting in the mode integral gives $k^2 F_{ij}(k) \propto q^{-2}$.

\subsection{Full Fisher integral}
The total Fisher matrix becomes
\begin{equation}
    F_{ab} = \int_{k_{\min}}^{k_{\max}} 
    \frac{k^2\,\mathrm{d}k}{(2\pi)^2}\,F_{ab}(k).
\end{equation}
We change to integration over $q$, define $\ell \equiv \ell_s / (a H)$ and obtain
\begin{align}
    F_{\alpha\alpha} &= 
    \frac{4V}{\ell^3(2\pi)^2}
    \int_{q_{\min}}^{q_{\max}}\!\!\mathrm{d}q\,q^2
    \left(\frac{1}{\alpha + q^2}\right)^2 \nonumber\\
    &=
    -\frac{4V(aH)^3}{\ell_s^3(2\pi)^2}
    \left[
    \frac{q}{2(\alpha + q^2)} + 
    \frac{\arctan(q/\sqrt{\alpha})}{2\sqrt{\alpha}}
    \right]_{q_{\min}}^{q_{\max}}.
\end{align}
The remaining integrals, $F_{\ell_s\ell_s}$ and $F_{\alpha\ell_s}$, are less tractable analytically but can be expanded in powers of $(\alpha - 1) \simeq 0.005$.  
To first order, we find:
\begin{widetext}
\begin{align}
    F_{\alpha\alpha} &\simeq
    \frac{4V(aH)^3}{\ell_s^3(2\pi)^2}
    \left[
    \frac{2q(q^2+1) - (\alpha-1)(q(q^2+1)+2q)}{4(q^2+1)^2}
    + \frac{3-\alpha}{4}\arctan(q)
    \right]_{q_{\min}}^{q_{\max}}, \\[4pt]
    F_{\ell_s\ell_s} &\simeq
    \frac{4V(aH)^3}{\ell_s^5(2\pi)^2}
    \left[
    2(1-\alpha)
    \left(
    \frac{3q^5 - 8q^3 - 3q}{48q^6 + 144q^4 + 144q^2 + 48}
    + \frac{\arctan(q)}{16}
    \right)
    \right]_{q_{\min}}^{q_{\max}}, \\[4pt]
    F_{\alpha\ell_s} &\simeq
    \frac{4V(aH)^3}{\ell_s^4(2\pi)^2}
    (2 - 2\alpha)
    \left[
    \frac{-5q^3 - 3q}{8q^4 + 16q^2 + 8}
    + \frac{3\arctan(q)}{8}
    \right]_{q_{\min}}^{q_{\max}}.
\end{align}
\end{widetext}
Neglecting coefficients, the leading-order scaling becomes
\begin{align}
    F_{\alpha\alpha} &\sim (3-\alpha)\frac{V(aH)^3}{\ell_s^3}
    \left[\frac{1}{q} + \arctan(q)\right]_{q_{\min}}^{q_{\max}}, \\[4pt]
    F_{\alpha\ell_s} &\sim (1-\alpha)\frac{V(aH)^3}{\ell_s^4}
    \left[\frac{1}{q} + \arctan(q)\right]_{q_{\min}}^{q_{\max}}, \\[4pt]
    F_{\ell_s\ell_s} &\sim (1-\alpha)\frac{V(aH)^3}{\ell_s^5}
    \left[\frac{1}{q} + \arctan(q)\right]_{q_{\min}}^{q_{\max}}.
\end{align}

Thus the uncertainty in $\ell_s$ scales as
\begin{equation}
    \sigma^2(\ell_s) \sim 
    \frac{F_{\alpha\alpha}}{\det\mathbf{F}}
    \sim \ell_s^5
    \quad
    \Rightarrow \quad
    \frac{\sigma(\ell_s)}{\ell_s} \sim \ell_s^{3/2}.
\end{equation}
Hence, as $\ell_s \rightarrow 0$, the parameter $\ell_s$ becomes increasingly well constrained.  
This scaling neglects the finite mode window $[q_{\min}, q_{\max}]$, and thus explains only the early-stage behavior observed numerically in Fig.~\ref{fig:SoundSpeed}. 

\newpage

\begin{figure*}
    \centering
    \includegraphics[width=1\linewidth]{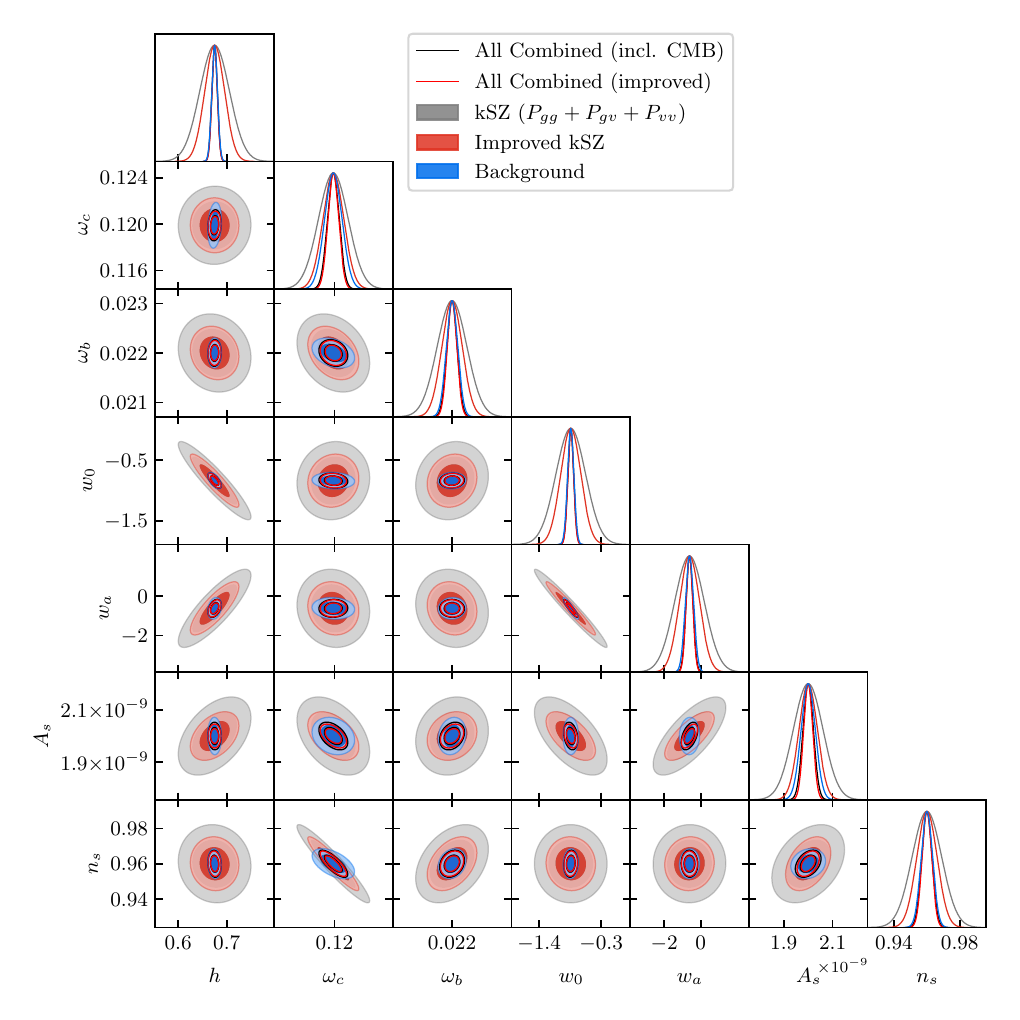}

\caption{Fisher matrices from the joint kSZ and background analysis discussed in the text. The \textit{improved} constraints correspond to CMB survey specifications with noise and beam parameters reduced by a factor of two relative to those in Eq.~(\ref{eq:CMBspec}).}
\end{figure*}

\end{document}